\documentclass{elsart}
\usepackage{amssymb}
\usepackage[dvips]{epsfig}

\begin{document}

\begin{frontmatter}

\title{Anderson localization problem: an exact solution for 2-D
anisotropic systems}

\author{V.N. Kuzovkov}

\address{Institute of Solid State Physics, University of
Latvia, \\ 8 Kengaraga Street, LV -- 1063 RIGA, Latvia}
\ead{kuzovkov@latnet.lv}

\author{ W.~von Niessen}

\address{Institut f\"ur Physikalische und Theoretische Chemie,
Technische Universit\"at Braunschweig, Hans-Sommer-Stra{\ss}e 10,
38106 Braunschweig, Germany}


\begin{abstract}
Our previous results [J.Phys.: Condens. Matter {\bf 14} (2002) 13777]
dealing with the analytical solution of the two-dimensional (2-D) Anderson localization
problem due to disorder is generalized for anisotropic systems (two different hopping
matrix elements in transverse directions). We discuss the mathematical nature
of the metal-insulator phase transition which occurs in the 2-D
case, in contrast to the 1-D case, where such a phase transition does not occur. In
anisotropic systems two localization lengths arise instead of one length only.
\end{abstract}


\end{frontmatter}



\section{Introduction}
Anderson localization \cite{Anderson58} remains one of the main
problems in the physics of disordered systems (see e.g. the review
articles \cite{Kramer,Janssen,Abrahams2}). In the series of our
previous papers \cite{Kuzovkov02,Kuzovkov04,Kuzovkov06} we
presented an exact analytic solution to this  problem. By the
exact solution we mean the calculation of the phase diagram for
the metal-insulator system. We have been able to solve the two
dimensional (2-D) problem \cite{Kuzovkov02}. We have shown then
that the phase of delocalized states exists for a non-interacting
electron system. The main aim of the paper \cite{Kuzovkov04} was
the generalization of the results to the case of higher
dimensional spaces (N-D). In paper \cite{Kuzovkov06} we discussed
the mathematical details of the new analytical approach for
calculating the phase-diagram. An exact solution is only possible
for the conventional Anderson model: the tight-binding
approximation with diagonal disorder, where on-site potentials are
independently and identically distributed.

It is well known that exact results in the field of phase
transitions (the metal-insulator transition is a particular case
of a phase transition) are exceedingly rare \cite{Baxter,Stanley}.
This is why any extension of the applicability range of analytical
methods in this field is of great interest. In this paper we
extend our approach \cite{Kuzovkov02,Kuzovkov04,Kuzovkov06,Reply}
for \textit{anisotropic} media
\cite{Li,Milde97,Milde00a,Milde00b}, but else remaining in the
framework of a conventional Anderson model.

Incorporation of an anisotropy into the tight-binding
approximation with diagonal disorder is also methodologically
valuable. Before the exact solution was obtained for $D > 1$
\cite{Kuzovkov02,Kuzovkov04,Kuzovkov06}, analytical methods
concentrated on the  $D=1$ case \cite{Lifshitz,Ishii}. However,
the specific topology of 1-D systems does not permit extension of
the results to higher dimensions. This is true in particular for
the Ising model, where an exact analytical solution for the 1-D
case has nothing to do with the 2-D solution obtained by Onsager
\cite{Baxter,Stanley}. As it is well known for the Anderson
localization problem, all states in the 1-D system are localized
(i.e. there is no metal-insulator transition). This is a
particular result of a general theory \cite{Stanley} that no phase
transitions are possible in 1-D systems with short-range
interactions. In fact, phase transitions (e.g. in the Ising model
\cite{Baxter,Stanley} or the Anderson localization problem
\cite{Kuzovkov02}) are observed only starting with $D=2$.

It should be realized that the approximate methods are also of a
limited use here. In particular, traditional perturbation theory
works perfectly in the 1-D case \cite{Lifshitz}, which is well
demonstrated by the analysis of the exact solution in Ref.
\cite{Kuzovkov02,Molinari}. Random potentials can be treated in
the 1-D Anderson localization problem as a small parameter, which
is used in series expansions of physical quantities. However, this
approach fails \cite{Stanley} for systems with phase transitions
in $D>1$, since physical quantities here are no longer described
by analytical functions and series expansions. The same is valid
for the Lyapunov exponent $\gamma$ (which is the inverse of the
localization length $\xi$ in the Anderson problem
\cite{Kuzovkov02}).

Of particular interest is the understanding of the mathematical
nature of the phase transition: how does the analytical character
of the exact solution for the 1-D problem changes to a
non-analytical character of the exact solution for the 2-D case?
For the \textit{discrete} spatial dimensions (1-D or 2-D) a simple
comparison of the two relevant solutions does not help us.
However, a ``\textit{continuous}'' treatment of the spatial
dimension for anisotropic systems, performed in this paper,
provides much more insight. We have put the word continuous in
quotation marks because we have in the present case of anisotropic
systems an interesting possibility of a transition from a 2-D
system to a 1-D system, by considering the limit $\kappa
\rightarrow 0$ for parameter of anisotropy $\kappa$.

The paper is organized as follows. In Section 2 we discuss the basic
equations for the isotropic problem \cite{Kuzovkov02}, which are
generalized there for the anisotropic case. Section 3 presents the
main results for the anisotropic problem. We demonstrate how the
study of the limiting case of a strong anisotropy permits us to
establish a relation between 1-D and 2-D cases.

\section{Recursion relation and the filter function}

\subsection{Isotropic system}

Let us start with the Schr\"odinger equation for the isotropic
system (the lattice constant and the hopping matrix element are set
equal to unity)
\begin{equation} \label{eq2}
\psi _{n+1,m}+\psi_{n-1,m}+\psi
_{n,m+1}+\psi_{n,m-1}=(E-\varepsilon _{n,m}) \psi _{n,m} .
\end{equation}
The on-site potentials $ \varepsilon_{n,m} $ are independently and
identically distributed with existing first two moments,
$\left\langle \varepsilon _{n,m}\right\rangle =0$ and
$\left\langle \varepsilon _{n,m}^2\right\rangle =\sigma ^2$.

Eq.(\ref{eq2}) can be written as the recursion
\begin{equation} \label{recursion 2}
\psi _{n+1,m}=-\varepsilon _{n,m} \psi _{n,m}-\psi_{n-1,m}+
\mathcal{L}\psi _{n,m} ,
\end{equation}
where the operator $\mathcal{L}$ acts on the index $m$ and is
defined by the relation
\begin{equation} \label{L}
\mathcal{L}\psi_{n,m} \equiv
E\psi_{n,m}-\sum_{m^{\prime}=\pm1}\psi_{n,m+m^{\prime}} .
\end{equation}
The standard initial (boundary) conditions are $\psi_{0,m}=0$ and
$\psi_{1,m}=\alpha_m$, respectively.

The existence of some fundamental 1-D numerical series $h_n$
($n=0,1,\dots,\infty$), the so-called system function or
\textit{filter}, was proved rigorously in Ref.
\cite{Kuzovkov02,Kuzovkov04,Kuzovkov06}.
A study of the
asymptotic behaviour of this series allows to define uniquely the
\textit{phase diagram} of the system. Namely, the series $h_n$ is
bounded
\begin{equation}\label{h_n1}
|h_n| < \infty ,
\end{equation}
for the delocalized states, but increases without bound
\begin{equation}\label{h_n2}
\lim_{n\rightarrow \infty} |h_n| \rightarrow \infty ,
\end{equation}
for the localized states. The actual numerical values are not
important in this respect. The determination of the existence
region of the bounded series, eq.(\ref{h_n1}), permits to find
also that for the delocalized states, i.e. to obtain the system's
phase diagram. In its turn, the asymptotics of the series
(\ref{h_n2}), $h_n \propto \exp(2\gamma n)$, for the localized
states allows us to extract the \textit{Lyapunov exponent}
$\gamma$, or the localization length $\xi=1/\gamma$.

It is more convenient to determine the phase diagram and the
Lyapunov exponent $\gamma$ by using the complex variable $z$
(Z-transform)
\begin{equation}\label{H_z}
H(z)=\sum_{n=0}^{\infty}\frac{h_n}{z^n} .
\end{equation}
In this case the physical problem of the localized/delocalized
states is reduced to the mathematical problem of the search for
the poles of the complex variable function $H(z)$
\cite{Kuzovkov02,Kuzovkov04,Kuzovkov06}, which is also called the
\textit{filter}. Eq.(\ref{h_n2})  means in this case that $H(z)$
has the pole at $z=\exp(2\gamma)$.

For the isotropic system the filter can be found exactly
\cite{Kuzovkov02,Kuzovkov04,Kuzovkov06}:
\begin{eqnarray}
 H^{-1}(z)=1-\frac{\sigma ^2}{2\pi}\frac{z+1}{z-1
}\int_{-\pi }^\pi \frac{dk}{w ^2-{\mathcal{L}}^2 (k)},\label{H}
\\ w ^2=\frac{(z+1)^2}z ,\label{w}
\end{eqnarray}
where
\begin{equation}\label{Lk}
\mathcal{L}(k)=E - 2\cos (k)
\end{equation}
corresponds to the operator $\mathcal{L}$ and arises due to use of
the Fourier transform.

Note that solution of the Schr\"odinger equation (\ref{recursion 2})
depends on chosen initial conditions, $\psi_{0,m}=0$ and
$\psi_{1,m}=\alpha_m$. In contrast, as follows from the definition,
eq.(\ref{H}), the filter $H(z)$  is independent of the initial
conditions (i.e. the field variables $\alpha_m$), and thus is
\textit{not} the functional of the wave function $ \psi _{n,m}$.
Following \cite{Kuzovkov02,Kuzovkov04,Kuzovkov06}, let us briefly
explain the mathematical nature of the filter $H(z)$. According to
the definition, eq.(\ref{L}), the operator $\mathcal{L}$ acts on the
wave functions. The expression eq.(\ref{H})  is a functional of
$\mathcal{L}$ and a disorder parameter  $\sigma$. The analysis
\cite{Kuzovkov02,Kuzovkov04,Kuzovkov06} shows that the function
$H(z)$ is also an \textit{operator} which acts, however, not on wave
functions but certain mathematical objects called \textit{signals}.
Only these signals are functionals of the wave functions.

The relation between the filter $H(z)$ and signals (input signal
$S^{(0)}(z)$ and output signal $S(z)$),
\begin{eqnarray}
S(z)=H(z)S^{(0)}(z), \\
 S^{(0)}(z)=\frac{1}{2\pi}\frac{z+1}{z-1
}\int_{-\pi }^\pi \frac{|\alpha(k)|^2 dk}{w ^2-{\mathcal{L}}^2
(k)},
\end{eqnarray}
has a certain similarity with the transition to the operator
formalism in quantum mechanics. Note that the boundary conditions
(field $\alpha_m$ or Fourier transform $\alpha(k)$) influence only
functions $S^{(0)}(z)$ which are independent of the parameter
$\sigma$. In other words, functions $S^{(0)}(z)$ correspond to
solutions in a \textit{completely ordered system}. Introduction of
disorder, $\sigma > 0$, transforms the initial solution
$S^{(0)}(z)$  into $S(z)$, where the operator $H(z)$ describes
this transformation.

The concept of the filter function is a general and abstract
description of the problem of localization. Instead of analyzing
wave functions, i.e. different signals, it is sufficient to analyse
properties of the fundamental \textit{localization operator} $H(z)$
by means of the theory for functions of complex variables. The
signal concept (input and output signals) and the filter function
are key elements in the theory of signals used earlier
\cite{Kuzovkov02,Kuzovkov04,Kuzovkov06}. In these papers the
connection between the Anderson localization problem and signal
theory is discussed and the most important concept of the proposed
method,  the filter $H(z)$,  is defined. The knowledge of papers
\cite{Kuzovkov02,Kuzovkov04,Kuzovkov06} presenting the main
formalism  is prerequisite for understanding the present paper.

Let us discuss equation (\ref{h_n2}). The localized states
correspond to values of $\gamma > 0$,  i.e. a divergence of the
filter function $h_n$ (or signals). This divergence is
mathematically similar to the divergence of averages in the
diffusion motion \cite{Kuzovkov06}. That is, transition from
delocalized to localized states can be treated as a
\textit{generalized diffusion} with a noise-induced first-order
phase transition. The generalized diffusion arises due to the
instability of a fundamental mode corresponding to averages of wave
functions (correlators and signals).

\subsection{Anisotropic system}\label{Aniz}

The Schr\"odinger equation for the 2-D anisotropic medium
\begin{equation} \label{eq2a}
t_1(\psi _{n+1,m}+\psi_{n-1,m})+t_2(\psi
_{n,m+1}+\psi_{n,m-1})=(E-\varepsilon _{n,m}) \psi _{n,m}
\end{equation}
contains two hopping matrix elements: $t_1,t_2$. Let us use the
units, where $\max \{t_1,t_2\}=1$, and $\min \{t_1,t_2\}=\kappa$
with $0 \leq \kappa \leq 1$. The parameter $\kappa$ characterizes
the system's anisotropy. For example, the bandwidth is $|E|\leq
E_{max}=2(1+\kappa)$.

Eq.(\ref{eq2a}) can also be written as a recursion relation. Let us
assume $\varepsilon^{\prime}_{n,m}=\varepsilon _{n,m}/t_1$ and
define the operator $\mathcal{L}^{\prime}$ through the relation
\begin{equation} \label{L2}
\mathcal{L}^{\prime}\psi_{n,m} \equiv
\frac{E}{t_1}\psi_{n,m}-\frac{t_2}{t_1}\sum_{m^{\prime}=\pm1}\psi_{n,m+m^{\prime}}
.
\end{equation}
As a result, we obtain the recursion
\begin{equation} \label{recursion 3}
\psi _{n+1,m}=-\varepsilon^{\prime}_{n,m} \psi
_{n,m}-\psi_{n-1,m}+ \mathcal{L}^{\prime}\psi_{n,m} ,
\end{equation}
functionally similar to eq.(\ref{recursion 2}). The equation for
the filter function $H(z)$ transforms, respectively, to
\begin{eqnarray}
 H^{-1}(z)=1-\frac{(\sigma^{\prime})^2}{2\pi} \frac{z+1}{ z-1
}\int_{-\pi }^\pi \frac{dk}{w ^2-({\mathcal{L}^{\prime}(k)})^2
},\label{H2}
\end{eqnarray}
where $(\sigma^{\prime})^2 =\left\langle
(\varepsilon^{\prime}_{n,m})^2\right\rangle =\sigma^2/t_1^2$,
\begin{equation}\label{Lk2}
\mathcal{L}^{\prime}(k)=\frac{E}{t_1} - \frac{t_2}{t_1}2\cos (k)
\end{equation}

In an anisotropic system there are two distinct directions. We give them
different names in order to distinguish them later. The choice
$t_1=1$, $t_2=\kappa$ corresponds to the direction of strong binding,
while $t_1=\kappa$, $t_2=1$ to the direction of weak binding.
Note that the former case (strong binding) should only be considered
for establishing the correspondence with the exact 1-D solution
\cite{Kuzovkov02,Molinari} in the limit $\kappa \rightarrow 0$.
Indeed, for $\kappa=0$ one gets $t_1=1$, $t_2=0$ and the operator
$\mathcal{L}^{\prime}=E$. Eq.(\ref{recursion 3}) transforms,
respectively, into the ensemble of similar equations characterized
by a neutral index $m$,
\begin{equation} \label{recursion 4}
\psi _{n+1,m}=-\varepsilon_{n,m} \psi _{n,m}-\psi_{n-1,m}+ E
\psi_{n,m} ,
\end{equation}
each of which is equivalent to the 1-D equation
\begin{equation} \label{recursion 5}
\psi _{n+1}=-\varepsilon_{n} \psi _{n}-\psi_{n-1}+ E \psi_{n} .
\end{equation}

\subsection{Useful relation}

Let us use the relation
\begin{equation}\label{eq103}
  \frac{1}{2\pi}\int_{-\pi}^{\pi} \frac{dk}{\zeta \pm 2\cos(k)} =
  \frac{1}{i \sqrt{4-\zeta^2}} ,
\end{equation}
which holds provided the imaginary part of the complex variable
$\zeta$ is non-negative, $Im \, \zeta \geq 0$ ($\zeta$ is defined
in the upper semi-plane).

The integrand (\ref{H2}) can be presented as
\begin{equation}\label{w01}
\frac{1}{w ^2-({\mathcal{L}^{\prime}(k)})^2
}=\frac{1}{2w}[\frac{1}{w-\mathcal{L}^{\prime}(k)}+\frac{1}{w+\mathcal{L}^{\prime}(k)}]
,
\end{equation}
and then eq. (\ref{eq103}) can be used, provided that the
parameter $w$ is also defined in the upper semi-plane, $Im \,  w
\geq 0$.

The transformation from the complex variable $z$ to $w$
(\textit{conformal mapping}) gives a parametric presentation of
the filter function $H(z)\equiv \mathcal{H}(w)$ \cite{Kuzovkov04}
convenient for analysis. Since the relation between the variables $z$ and
$w$ is \textit{nonlinear}, this relation is not \textit{unique}.

\subsection{Conform transformation and phase index $P$}

If the complex variable $w=u+iv$ is defined in the upper
semi-plane, $v\geq 0$, the relation
\begin{equation}\label{wz}
w=P(z^{1/2}+z^{-1/2})
\end{equation}
is valid \cite{Kuzovkov02}, where the parameter $P=\pm 1$. As
shown below, $P$ serves as an index numbering different-type
solutions: localized ($P=+1$) and delocalized ($P=-1$), i.e. it
characterizes the phase state of a system.

For $P=+1$ eq.(\ref{wz}) characterizes the conform transformation
of the outer region of the unit circle, $|z|\geq 1$, onto the
upper $w$-semi-plane. Similarly, for $P=-1$ the conform
transformation of the inner part of the unit circle, $|z|\leq 1$,
occurs.

The inverse transformation is described by the relations
\begin{eqnarray}\label{zw}
z=-1+\frac{w^2}{2}+P\frac{w i}{2}\sqrt{4-w^2} ,
\end{eqnarray}
provided
\begin{eqnarray}\label{zw2}
\frac{z+1}{z-1}=P\frac{w}{i \sqrt{4-w^2}} .
\end{eqnarray}

\subsection{Anderson localization and first-order phase transition }

As was shown \cite{Kuzovkov02,Kuzovkov04}, for arbitrary dimension
$D$ the phase of localized states (the solution with $P=+1$) is found
for an arbitrary disorder $\sigma$. In contrast, the
delocalized phase ($P=-1$) arises only for $D \geq 2$. It exists
within the limited energy band and below the critical disorder
parameter $\sigma$.

A comparison of the 1-D and 2-D cases shows their qualitative
differences. In 1-D even infinitesimal disorder destroys the
localized states; their existence range is degenerate into a
single point $\sigma \equiv 0$ \cite{Kuzovkov02,Molinari}.
Similarly to many other problems of phase transitions
\cite{Baxter}, the (metal-insulator) phase transition in 1-D is
impossible. In this respect the delocalized states at $\sigma
\equiv 0$, which are unstable under any perturbation, cannot be
treated as a special phase. It is more reasonable to treat these
states as the limiting case of the localized states which only
exist for 1-D. As a consequence, the physical characteristics of the
single-phase system such as the Lyapunov exponents,
$\gamma=\gamma(\sigma,E)$, are \textit{analytical functions} of
the energy $E$ and the disorder parameter $\sigma$. Moreover, the
perturbation theory also applies here: $\gamma=\gamma(\sigma,E)$ can
be expanded in a series of a small parameter $\sigma^2$
\cite{Kuzovkov02}. The 1-D system is characterized by a
\textit{continuum} transition due to the lack of disorder, $\sigma
\rightarrow 0$. The limiting value $\lim_{\sigma \rightarrow
0}\gamma(\sigma,E)=\gamma(0,E)$  has obviously to coincide with
that for the delocalized states, $\gamma=0$. In other words,
\begin{equation}\label{gam}
\gamma(0,E)=0
\end{equation}
holds.

The 2-D case differs \textit{qualitatively} from the 1-D case. First
of all, in the existence range of the second solution ($P=-1$) the
first solution ($P=+1$) is also always present. In other words,
two phases \textit{co-exist} \cite{Kuzovkov02,Kuzovkov04}.
Therefore, the metal-insulator transition has to be interpreted as a
\textit{first-order} phase transition. Two fundamental points
should be stressed here. (i) In any system with dimension larger
or equal to two and with a phase transition, the
physical properties are no longer analytical functions of their
variables. That is, such properties as the Lyapunov exponents
$\gamma(\sigma,E)$ can manifest peculiarities, e.g. step-like
changes, and thus the expansion in \textit{small parameter}
$\sigma^2$ no longer holds here \cite{Kuzovkov02}.  (ii) The
first-order phase transitions reveal additional peculiarities. The
coexisting phases (in our case, the delocalized states with
$\gamma=0$ and localized states with $\gamma(\sigma,E)\neq 0$)
have nothing in common, and thus there is no reason to expect that
eq.(\ref{gam}) holds. That is, in a system with first-order
phase transition
\begin{equation}\label{gam2}
\gamma(0,E)\neq 0 ,
\end{equation}
which illustrates once more the above-discussed non-analytical
behaviour of the physical properties.

As was recently noted \cite{Kuzovkov02}, $\gamma(\sigma,E)$ in the
Anderson localization problem can be treated as the
\textit{long-range} order parameter. Indeed, the two different
phases reveal different $\gamma$ parameters, $\gamma \equiv 0$ and
$\gamma \neq 0$, respectively. The first-order transition is
always characterized by the step-like change of the long-range
parameter \cite{Stanley}. In this respect, eq.(\ref{gam2}) being
the particular case of a more general relation $\gamma \neq 0$ for
the phase of delocalized states, seems quite self-evident, despite
possible exceptions discussed below.

Therefore, the analysis of the limiting transition to 1-D in the
anisotropic system permits, on the one hand, to understand the
mechanism for the change from non-analytical to analytical solutions
for the physical properties,
and, on the other hand, gives an additional check of our
results \cite{Kuzovkov02,Kuzovkov04} since the difference between
eqs.(\ref{gam}) and (\ref{gam2}) is nontrivial.

\section{Results}

\subsection{General remarks}

Using the above-obtained relations, one gets the following parametric
presentation of the filter function
\begin{equation}\label{Hw}
\frac{1}{\mathcal{H}(w)}=1+P\frac{\sigma^2}{2t_1t_2\sqrt{4-w^2}}\sum_{\nu=\pm
1}\frac{1}{\sqrt{4-Q^2_{\nu}(w,E)}} ,
\end{equation}
where
\begin{equation}\label{Q}
Q_{\nu}(w,E)=\frac{t_1}{t_2}w+\nu \frac{E}{t_2} .
\end{equation}
Let us consider now the properties of the filter function
$\mathcal{H}(w)=\mathcal{H}_{-}(w)$ at $P=-1$. As was shown in
Ref.\cite{Kuzovkov02,Kuzovkov04}, the filter analysis permits to
obtain the phase diagram, i.e. to determine the existence range of
the delocalized states. For this one has to determine the poles of $H(z)$,
i.e. to find the roots of the equation $H^{-1}(z)=0$. In the
parametric presentation for $P=-1$  we seek roots of
\begin{equation}\label{Hw-}
\mathcal{H}^{-1}_{-}(w)=0 ,
\end{equation}
or
\begin{equation}\label{Hw--}
\frac{\sigma^2}{2t_1t_2\sqrt{4-w^2}}\sum_{\nu=\pm
1}\frac{1}{\sqrt{4-Q^2_{\nu}(w,E)}} =1 .
\end{equation}
The peculiarity of the case $P=-1$ \cite{Kuzovkov02,Kuzovkov04} is
that its physical interpretation is possible either in the
\textit{absence} of the poles (which is the case for $D > 2$
\cite{Kuzovkov04}), or if the poles lie on the unit circle $|z|=1$
(\textit{marginal stability}), i.e.
\begin{equation}\label{zphi}
z=\exp(\pm i 2\varphi)
\end{equation}
(pairs of complex values), $\varphi \in(0,\pi)$. The letter case
applies to 2-D isotropic systems \cite{Kuzovkov02}. It can also be
shown using the method in ref. \cite{Kuzovkov04} that taking into
account the anisotropy does not affect eq.(\ref{zphi}) (we omit here
the proof). As a result, in the general case of an anisotropic
system it follows from eq.(\ref{wz}) that
\begin{equation}\label{wz-}
w=-2\cos(\varphi) ,
\end{equation}
i.e. the $w$ parameter is \textit{real}. This obviously simplifies the
solution of eq.(\ref{Hw--}).

The equation
\begin{equation}\label{diag-}
\mathcal{H}^{-1}_{-}(w=0)=0
\end{equation}
determines the threshold disorder $\sigma=\sigma_0(E)$ which
defines the limits of existence of the delocalized phase
\cite{Kuzovkov02,Kuzovkov04}.

It is easy to conclude that the real roots $w$ of eq.(\ref{Hw-})
satisfy the condition $|Q_{\nu}(w,E)|\leq 2$, which is equivalent to
\begin{equation}\label{modw+}
t_1|w|+|E| \leq 2 t_2 .
\end{equation}
Simultaneously $0\leq |w|\leq 2$ has to hold.

Consider now the filter function $\mathcal{H}_{+}(w)$ for $P=+1$.
This function has always the only pole \cite{Kuzovkov02}
\begin{equation}\label{zgamma}
z=\exp(2\gamma) ,
\end{equation}
which lies on the real axis with $\gamma \geq 0$, where $\gamma$
is interpreted as the Lyapunov exponent. In the parametric
presentation the parameter $w$ is also real,
\begin{equation}\label{wz+}
w=2\cosh(\gamma) .
\end{equation}
To find the Lyapunov exponent, one has to determine the root of
the equation
\begin{equation}\label{diag+}
\mathcal{H}^{-1}_{+}(w)=0 ,
\end{equation}
or
\begin{equation}\label{Hw+}
\frac{\sigma^2}{2t_1t_2\sqrt{w^2-4}}[\frac{1}{\sqrt{Q^2_{+}(w,E)-4}}+\frac{1}{\sqrt{Q^2_{-}(w,E)-4}}]=1
.
\end{equation}
The parameter $w$ sought for satisfies the natural condition of
positive arguments of all roots of eq.(\ref{Hw+}), so the simple
condition
\begin{equation}\label{ww+}
t_1 w -|E| \geq 2 t_2
\end{equation}
has to be fulfilled, along with $w \geq 2$.

\subsection{Isotropic system}

Let us summarize for further analysis the main results for an
isotropic system.

\subsubsection{The case of $P=-1$ (delocalized states).}

In an isotropic system the real roots $w$ have to fulfill
$0\leq |w|\leq 2$ \cite{Kuzovkov02}. This is possible only in the
limited energy range $|E| \leq E_0=2$, otherwise one obtains
complex roots with an unclear interpretation. That is, the
delocalized states exist only within a certain energy range around
the zone center. In this energy range an increase of the disorder
parameter $\sigma$ shifts the poles in the parametric presentation
from $|w|=w_0=2-|E|\leq 2$ towards the limiting value of $w=0$.
For further disorder increases, the real roots disappear and thus
the formal solutions of eq.(\ref{Hw-}) permit no physical
interpretation.


Using eq.(\ref{diag-}) at $t_1=t_2=1$,  one gets \cite{Kuzovkov02}
\begin{eqnarray}\label{sigma_0}
\sigma_0(E)=2 (1-E^2/E^2_0)^{1/4} .
\end{eqnarray}

\subsubsection{The case $P=+1$ (localized states).}

Eq.(\ref{ww+}) determines the existence of real roots $w \geq
u_0=2+|E|$  of eq.(\ref{Hw+}) \cite{Kuzovkov02,Kuzovkov04}. The
eq.(\ref{wz+}) holds here with $\gamma=\gamma(\sigma,E)$. In the
parametric presentation the value of $w=u_0$ corresponds to the
absence of disorder, $\sigma=0$. In this case
\begin{equation}\label{gammaE}
\gamma(0,E)=\sinh^{-1}(1+\frac{|E|}{2}) .
\end{equation}
This result confirms eq.(\ref{gam2}). The continuity,
eq.(\ref{gam}), is found only at the band center, $E=0$, when
$\gamma(0,0)=0$. In this case
one obtains for the Lyapunov exponent
\begin{equation}\label{gamma0}
\gamma(\sigma,0)=\sinh^{-1}(\frac{\sigma}{2}) .
\end{equation}
E.g. for  $\sigma \rightarrow 0$ one has from eq.(\ref{gamma0})
$\gamma(\sigma,0) \propto \sigma$. It follows from this that the
function $\gamma(\sigma,E)$ cannot be represented as a series in
powers of $\sigma^2$, i.e. perturbation theory is not applicable
to the 2-D Anderson problem.

\subsection{Strong binding direction}

Let us now assume that $t_1=1$, $t_2=\kappa$. As was mentioned in
section \ref{Aniz}, this case in the limit of $\kappa \rightarrow
0$ corresponds to the transition to the 1-D case.

\subsubsection{The case $P=-1$ (delocalized states).}

In this case eq.(\ref{modw+}) transforms into
\begin{equation}\label{modw1}
|w|+|E| \leq 2\kappa .
\end{equation}
Along with the relation $0\leq |w|\leq 2$ one obtains again that the real
roots exist in the energy range of $|E| \leq E_1=2\kappa$.
A detailed study shows that this situation is similar to that for
the isotropic system, with a marginal stability of the filter function.


The threshold disorder value reads now
\begin{eqnarray}\label{sigma_1}
\sigma_1(E)=2 \sqrt{\kappa} (1-E^2/E^2_1)^{1/4} .
\end{eqnarray}
Therefore, on can see a monotonic decrease of both energy
half-width $E_1 \propto \kappa$ (where delocalized states exist)
and the threshold disorder $\sigma_1(E)\propto \sqrt{\kappa}$ in
the limit $\kappa \rightarrow 0$ in the strong binding direction.
For $\kappa=0$ the phase of delocalized states loses its existence
range, which corresponds to the well-known 1-D result.

\subsubsection{The case of  $P=+1$ (localized states).}

Eq.(\ref{ww+}) transforms into $w \geq 2\kappa+|E|$. Together with
$w \geq 2$ one gets
\begin{equation}\label{wa}
w \geq u_0=\max \{2, 2\kappa+|E| \} .
\end{equation}
For the lack of disorder, $\sigma \rightarrow 0$, $w \rightarrow
u_0$.

It is easy to find that at $|E|\leq E_2=2(1-\kappa)$ the parameter
$u_0=2$. Respectively, in this energy range for $\sigma=0$ the
Lyapunov exponent $\gamma(0,E)=0$.  At small $\sigma$ the Lyapunov
exponent is proportional to $\sigma^2$: $\gamma(\sigma,E) \sim
\beta \sigma^2$. Analysis of eq.(\ref{diag+}) yields
\begin{equation}\label{beta}
\beta(E,\kappa) = \frac{1}{4}
[\frac{1}{\sqrt{(E_{max}+E)(E_2+E)}}+
\frac{1}{\sqrt{(E_{max}-E)(E_2-E)}}]
\end{equation}

In the 1-D limit, $\kappa \rightarrow 0$, one gets $E_{max}=2$,
$E_2=2$, $|E|\leq 2$, respectively
\begin{equation}\label{beta0}
\beta(E,0)=\frac{1}{4-E^2} .
\end{equation}
As one can expect, this corresponds to the exact 1-D result
\cite{Molinari} (in the limit of small $\sigma$):
\begin{equation}
\gamma(\sigma,E)\sim \sigma^2 /(4-E^2) .
\end{equation}

In the energy range $E_2 \leq |E|\leq E_{max}$ $u_0=2\kappa +|E|
\geq 2$. Respectively, the value
\begin{equation}\label{gamma3}
\gamma(0,E)=\cosh^{-1}(\kappa +\frac{|E|}{2})
\end{equation}
is nonzero.


Therefore, in the \textit{isotropic} system with $\kappa=1$ due to
the presence of the phase transition eq.(\ref{gammaE}) confirms
completely the general conclusion, eq. (\ref{gam2}). The only
exception is the band center $E=0$, where $\gamma(0,0)=0$ holds,
and there is a correspondence with the main relation for the 1-D
system - eq.(\ref{gam}). As \textit{anisotropy} arises, $\kappa
> 0$, this point moves into the energy range $|E|\leq E_2$, where
eq.(\ref{gam}) holds. Moreover, within this range
$\gamma(\sigma,E)$ can be expanded into a series in a small
parameter $\sigma^2$, similarly to the 1-D case. Outside this
energy range eq.(\ref{gam2}) remains valid. As we have
demonstrated, at $\kappa=0$ our equations transform into those
obtained earlier for the 1-D system \cite{Kuzovkov02,Molinari}.

\subsection{Weak binding direction}

Let us assume now $t_1=\kappa$, $t_2=1$.

\subsubsection{The case $P=-1$ (delocalized states).}

In this case the following relations
\begin{equation}\label{modw2}
\kappa |w|+|E| \leq 2
\end{equation}
and $0\leq |w|\leq 2$ hold.

Real roots of equation $\mathcal{H}^{-1}_{-}(w)=0$ exist in the
parameter range $0 \leq |w|\leq \min \{2, (2-|E|)/\kappa \}$ for
energy $|E| \leq E_0=2$. The threshold disorder magnitude, where
the delocalized states still exist, remains to be defined by
eq.(\ref{diag-}):
\begin{eqnarray}\label{sigma_2}
\sigma_2(E)=2 \sqrt{\kappa} (1-E^2/E^2_0)^{1/4} .
\end{eqnarray}


\subsubsection{The case $P=+1$ (localized states)}

Eq.(\ref{ww+}) corresponds to
\begin{equation}\label{wa++}
w \geq u_0= (2+|E|)/\kappa ,
\end{equation}
whereas the second condition $w \geq 2$ is fulfilled
automatically. As $\sigma \rightarrow 0$,  $w \rightarrow u_0>2$
holds. Therefore eq.(\ref{gam2}) holds in the whole energy range.


\subsection{Phase diagram}

The delocalized state in the anisotropic system can be discussed
only provided delocalization occurs simultaneously in two
dimension. It is easy to see that decisive point here are relations
obtained for the strong binding direction. Therefore, the phase of
delocalized states exists in the range $|E|\leq E_1=2\kappa$,
provided for a given energy the disorder does not exceed the
critical value, eq.(\ref{sigma_1}). A decrease of the anisotropy
parameter $\kappa$ leads to a monotonic decrease of both the
energy range and the threshold disorder.

\section{Conclusions}

The phase of localized states in the anisotropic system can be
characterized by \textit{two} Lyapunov exponents
$\gamma(\sigma,E)$ and, respectively, two localization lengths.
The Lyapunov exponent in the strong binding direction is always
smaller then that in the weak binding direction. In other words,
the localization length in the strong binding direction is always
larger then that in the perpendicular direction. These two
localization lengths reveal different asymptotic behaviour as
$\sigma \rightarrow 0$. In the limit of strong anisotropy, $\kappa
\rightarrow 0$, only the localization length in the strong binding
direction serves as an analogue to the single localization length
in the 1-D system.

\ack{V.N.K. gratefully acknowledges the support of the Deutsche
Forschungsgemeinschaft. The authors are indebted to E. Kotomin for
detailed discussions and comments on the paper.}


\end{document}